\RequirePackage[2020-02-02]{latexrelease}
\documentclass[10pt]{revtex4}
\usepackage{graphicx}
\usepackage{amsmath,amssymb}
\usepackage{mathtools}

\begin{document}

\title{Can entanglement be mediated by a Koopmanian system?}
\author{Chiara Marletto and Vlatko Vedral}
\affiliation{Clarendon Laboratory, University of Oxford, Parks Road, Oxford OX1 3PU, United Kingdom}

\begin{abstract}
We present a method for coupling a Koopmanian classical system to two quantum bits to mediate an interaction between them. We then prove that the resulting dynamics can never lead to entanglement between the two qubits. Even though the total system of two qubits and the Koopmanian classical system are described with the full quantum formalism, we show that their composite system violates exact conservation laws as expected for a hybrid quantum-classical system. We finally discuss the implications for semi-classical treatments of quantum gravity. 
\end{abstract}

\maketitle

We analyse a model of classical dynamics within the formalism of quantum mechanics, based on the Koopmanian approach \cite{Koopman}. The reason for doing this is to investigate hybrid quantum-classical systems using the same joint formalism: many proposals have been put forward in the literature, \cite{Koopman, Sudarshan, KIEREG, HARE08}, but their validity is still hotly debated. Despite the fact that there are claims that such models can successfully capture semi-classical gravity, we show that this is not the case. This result is inline with arguments that date back to B. DeWitt's seminal work, \cite{DEW, DeW-Book, MV-njp, MV-njpq}, proving that if a sector of the universe is quantum and can couple, under a number of plausible assumptions, with another sector, the latter cannot be purely classical.

We start with expressing the classicality condition for a system $S$, as the constraints that all the operators representing physical variables of $S$ must commute with each other.

We start with $S$ being a free classical particle whose canonical variables are $x$ and $p$. By definition of being classical, $x$ and $p$ commute. It is therefore possible to encode them into degrees of freedom pertaining to two different quantum systems. One such way is to assign the position of the classical particle to the position operator of the first quantum subsystem, $x=\hat x_1$, while at the same time assigning the momentum to the momentum operator of the second quantum subsystem,  $p=\hat p_2$. Because they belong to different subsystems, $[\hat x_1, \hat p_2] =0$ just as one has in classical mechanics. Also, the state of the two could be a simultaneous eigenstate of both $\hat x_1$ and $\hat p_2$, say 
\[
\psi (x_1,p_2,t=0) = \delta (x_1-\bar x_0)\delta (p_2-\bar p_0),
\] 
which represents a classical particle starting at the position $\bar x_0$ with the momentum $\bar p_0$. The other two variables, $x_2$ and $p_1$ are assumed to be hidden. They are never measurable, an assumption that we will revisit later in our paper. 

Hamilton's formulation of Newton's laws relies on the use of the canonically conjugate position $x$ and momentum $p$ and the equations of motion follow from the Hamiltonian $H=H(x,p)$:
\[
\frac{\partial x}{\partial t} = \frac{\partial H}{\partial p} \;\;\;\;\; \frac{\partial p}{\partial t} =- \frac{\partial H}{\partial x}
\]
In the case of a free particle, $H=p^2/2m$ and the above equations give us $\dot x = p/m$ and $\dot p = 0$, where $m$ is the mass of the particle. These are easily solved and lead to the usual equation for inertial motion $x(t) = x_0+pt/m$. 

What kind of quantum Hamiltonian of the two quantum subsystems describes the classical dynamics of the free particle? The quantum wavefunction of the two free particles must evolve in time into: $\psi (x_1,p_2,t) = \delta (x_1-\bar x_0 - \bar p_0 t/m)\delta (p_2-\bar p_0)$. Therefore, it is straightforward to conjecture that the Hamiltonian is:
\[
\hat H = \frac{\hat p_1 \hat p_2}{m}\;.
\]
We note that to reproduce the dynamics of the classical Hamiltonian of a free particle, $H=p^2/2m$, the factor of $1/2$ must be omitted from the quantum Hamiltonian. Note that there is no spreading of the wave-packet of a free particle characteristic of quantum dynamics. The reason is that the position of the particle always remains sharp as is the case for classical evolution. 

We now extend this treatment to include a force acting on the particle. It will suffice to consider a simple particle undergoing harmonic motion. In this case, the classical Hamiltonian becomes $H=p^2/2m + kx^2/2$, where $k$ is the spring constant. Likewise, the quantum Hamiltonian is 
\[
\hat H = k\hat x_1\hat x_2 
\]
and, note again, that the factor $1/2$ needs to be omitted. In order to obtain the dynamics, the easiest approach is to construct the unitary and apply it to the initial state
\[
\psi (x_1,p_2,t) = e^{-i (\hat p_1 \hat p_2/m + k\hat x_1\hat x_2)t} \psi (x_1,p_2,t=0)\; .
\]
Because the two terms in the Hamiltonian do not commute there are many ways to handle this, the easiest perhaps being a perturbative expansion. A calculation for the time displacement $\delta t$ leads to the state
\[
\psi (x_1,p_2,\delta t) = \delta (x_1-\bar x_0 - \bar p_0 \delta t/m)\delta (p_2-\bar p_0 - k \bar x_0 t)  
\]
from which it is clear that $\delta x = p/m \delta t$ and $\delta p = -kx\delta t$ which is the equation of motion for a particle in a harmonic potential (where, as explained the coordinate $x$ is the position of the first subsystem and $p$ is the momentum of the second subsystem). 

The quantum implementation of classical dynamics is most transparent in the Heisenberg picture. Consider a Hamiltonian of the form 
\[
H= \frac{\hat p_1 \hat p_2}{m} + V^\prime (\hat x_1) \hat x_2 \; ,
\]
where $V^\prime (\hat x_1)$ is the spatial derivative of the potential the particle is subjected to. The relevant Heisenberg equations of motions are now $\partial \hat x_1/\partial t = \hat p_2/m$ and $\partial \hat p_2/\partial t = V^\prime (\hat x_1)$. Since $\hat x_1$ and $\hat p_2$ commute, they behave like c-numbers and these two equations together constitute Newtonian dynamics. Note that when $V^\prime (\hat x_1) = 0$, so that no force acts on the particle, there is no resulting spread of the wavepacket. This is exactly as we would expect from classical trajectories. 

For completeness, we point out that the Galilean transformations are in this picture implemented by the unitary transformation of the form
\[
U(a,mv) = e^{i(a\hat p_1 -mv \hat x_2)}
\]
where $a$ is the spatial displacement and $mv$ is the boost. In the Heisenberg picture, this leads to the transformations
\begin{eqnarray}
U\hat x_1 U^{\dagger} & = & \hat x_1 + a \\
U\hat p_2 U^{\dagger} & = & \hat p_2 - mv 
\end{eqnarray}
which is exactly as expected. Unlike in quantum mechanics, here we do not obtain the extra phase due to the non-commutativity of $x$ and $p$, and this is, again, exactly as expected from classical dynamics. 

Let us now couple this classical system to two quantum bits so that the resulting interaction Hamiltonian becomes $\lambda \hat x_1 (\sigma^1_z + \sigma^2_z)$, where the $\sigma^1_z$ and $\sigma^2_z$ are the usual Pauli $z$ operators on the qubits $1$ and $2$ respectively. We will show that this Koopmanian mediator between the qubits cannot lead to entanglement generation between the qubits. We choose a simple total Hamiltonian to illustrate the point; however, it will become transparent that any other choice must lead to the same conclusion:
\[
\hat H = \frac{\hat p_1 \hat p_2}{m} + \lambda \hat x_1 (\sigma^1_z + \sigma^2_z)
\]
where $\lambda$ is the interaction strength. Note that we have omitted the potential in the Koopmanian part since it does not play any additional role as far as the interaction between the qubits is concerned. The dynamics is again seen from exponentiating the Hamiltonian which can now easily be done exactly
\[
e^{-i (\hat p_1 \hat p_2/m + \lambda \hat x_1 (\sigma^1_z + \sigma^2_z))t} = e^{-i \hat p_1 \hat p_2 t/m} e^{-i \lambda \hat x_1 (\sigma^1_z + \sigma^2_z)t} e^{-i \lambda \hat p_2/m (\sigma^1_z + \sigma^2_z)t}\; .
\]
The first term on the RHS is irrelevant as far as the coupling between the qubits. The second and the third terms both have the form $U_1\otimes U_2$ where the first unitary only acts on the qubit one and the second only on the qubit two. Therefore, the two qubits can never become entangled. It is clear that the addition of a potential to the Koopmanian system does not change this fact; also, all other individual couplings of qubits to the Koopmanian system will result in products of unitaries and will therefore not be able to generate entanglement. This is because of the classical condition (implemented in our model by the fact that the observables of the two different systems commute). 

To get entanglement, one would have to use a Hamiltonian that violates the classical condition, engaging at least two degrees of freedom that do not commute. For instance, adding a term of the kind:
\[
\hat H_{ENT} =  \alpha (\hat x_1\hat p_2 + \hat p_1 \hat x_2) + \lambda_1\sigma^1_x x_1 +\lambda_2 \sigma^2_z p_2 \;,
\]
where $\alpha, \lambda_1$ and $\lambda_2$ are some coupling constants, would allow one to entangle the two qubits, because the mediator is now engaged with the non-commuting variables $\hat x_1, \hat p_1$ and $\hat x_2, \hat p_2$.

This is an illustration of the fact that no classical mediator can entangle quantum systems, as proven by us in several papers, \cite{MAVE, MAVE2,MV-PRD}, using the general methodology of the Constructor Theory, \cite{DEUMA} (the less general argument based on LOCC is presented here \cite{SOUG}). Our main motivation was to show that a semi-classical account of gravity cannot account for entanglement generation between material qubits, as we have now demonstrated.

Now we would like to make an additional point. The Koopmanian treatment outlined here, however, gives the impression that the semi-classical account to couple the quantum and classical sectors is fully consistent, as it uses the full machinery of Hilbert spaces and the fact that the degrees of freedom of separate subsystems commute with one another. But this is not the case, as we shall now argue. The trouble with semi-classical models like the one presented here is that there cannot be any ``back-reaction" from the quantum on the classical system, as many have already noted in the literature \cite{Terno1, MV-AB, Marconato}. To have this feature, one would need to engage the hidden degrees of freedom in the Koopman model, namely the variables $x_2$ and $p_1$, as in the interaction term $\hat H_{ENT}$. If one were to do that, one would obtain a fully quantum system of two coupled subsystems with complementary variables $(\hat x_1, \hat p_1)$ and $(\hat x_2, \hat p_2)$. Imagine for instance that a conservation law applies to the composite system of one of the qubits and the classical system $S$: for instance, let's assume that the quantity $\hat C=\sigma^1_z+x_1$ is conserved. This means that the only allowed unitaries on $Q_1$ and $S$ jointly must commute with $\hat C$: $[U_{Q_1S}, \hat C]=0$. It is easy to check that the only unitaries satisfying this exact conservation law are generated by Hamiltonian of the form given by $H$ above, which can only lead to trivial local evolutions of the qubit sector. Instead, to induce on the qubit a transition from e.g. an eigenstate of the locally conserved quantity $\sigma^1_z$ to a different state that does not commute with $\sigma^1_z$, a Hamiltonian of the form given by $H_{ENT}$ would have to be adopted. This type of Hamiltonians violate the classicality condition, as expected considering numerous arguments in the literature \cite{DIP, FENG, Terno1, Terno}. 

In the concrete case of semi-classical gravity, if a mass is put in a spatial superposition of two different locations, then in each of these locations there should be a back-action on the underlying gravitational field if the field was quantum. However, if the field remains classical, this kind of back-action is impossible. The field must remain in a classically well-defined state, where all variables are simultaneously sharp. Therefore the quantum uncertainty that exists in the matter degrees of freedom does not get imprinted into the classical field. 

Finally, explaining even the ordinary static interactions between charges by using the field as a mediator is problematic here. The reason is that in the model where the Hamiltonian contains terms $x^2 + x(\sigma^1_z + \sigma^2_z)$, the coupling between $1$ and $2$ is seen by rewriting the Hamiltonian as $(x + (\sigma^1_z + \sigma^2_z))^2 - \sigma^1_z \sigma^2_z$ and the last term is the interaction between the subsystems $1$ and $2$ induced by each coupling individually to the coordinate $x$ of the field. In the Koopmanian models, this cannot happen because the Hamiltonian is linear in $x_1$ and $p_2$ and this is necessary in order for the quantum system to fully reproduce the classical behaviour. 

One last consideration is about claims that are often made regarding those arguments that, in line with what we have just discussed, show that hybrid classical-quantum models are inconsistent. Those arguments have their roots in a seminal work by DeWitt, \cite{DEW} and they often assume that the dynamics is deterministic. For this reason, some critics have dismissed them as they prima facie do not apply to stochastic models (see e.g. \cite{OPP}). Those criticisms however are erroneous because those arguments assume determinism for the evolution of dynamical variables, not all of which may be directly observable. All stochastic models that are suitable descriptions of physical reality should admit a deterministic model, which adopts dynamical degrees of freedom that are not directly observable to provide an underlying mechanism for the stochastic dynamics. Hence, the arguments in question can be applied to such extended models and are valid to rule out the semiclassical stochastic descriptions, too. This observation is a generalisation of the fact that CPTP (completely positive trace-preserving) maps can always be expressed, mathematically, by a class of unitaries acting on a larger Hilbert space. Such space may or may not be physical, but the results proven at the higher level must be then applicable to the reduced stochastic map itself. If one insists on stochastic models being fundamental, then one has to give up all the exact symmetries that hold for the quantum sector of the hybrid system. 

It is worth pointing out that in arguments such as those underlying the gravitationally induced entanglement \cite{MAVE,MV-PRD} and the related recent non-classicality witnesses \cite{MV-njpq, DIP}, one does not need to make that assumption of deterministic evolution; instead, one only relies on performing a sufficient number of measurements on the probe's state. This makes the witnesses even more compelling in ruling out semiclassical models.

To summarise: we have presented a simple model where a fully classical system is coupled to two qubits. We have first showed that it cannot generate entanglement between the qubits; and in order to do that, it must engage non-commuting degrees of freedom. We have also argued that the model itself is inconsistent as the classicality condition does not allow for back reaction of the quantum system onto the classical system. This once more shows that hybrid semiclassical models are useful approximations in some context but cannot provide a fundamental model for physical reality. 

\textit{Acknowledgments}: This research was made possible by the generous support of the Gordon and Betty Moore Foundation, the Eutopia Foundation, and the John Templeton Foundation, as part of The Quantum Information Structure of Spacetime (QISS) Project (qiss.fr). The opinions expressed in this publication are those of the authors and do not necessarily reflect the views of the John Templeton Foundation.

\end{document}